\documentclass[10pt,prl,aps,showpacs,twocolumn,preprintnumbers,amsmath,amssymb,superscriptaddress]{revtex4-2}
\usepackage{graphicx}    
\usepackage{latexsym}    
\usepackage{color}       
\usepackage{dcolumn} 
\usepackage{bm}          
\usepackage{amssymb}   
\expandafter\let\csname equation*\endcsname\relax
\expandafter\let\csname endequation*\endcsname\relax
\usepackage{amsmath}
\usepackage{mathtools}  
\usepackage{bbm}
\usepackage{graphics}
\usepackage{epsfig}
\usepackage{subfigure}
\usepackage{nicefrac}
\usepackage{verbatim}
\usepackage{enumerate}
\usepackage{footnote}
\usepackage{booktabs}
\usepackage{siunitx}
\usepackage{xcolor}
\renewcommand{\arraystretch}{1.8}
\usepackage[colorlinks=true,allcolors=blue]{hyperref}  
\usepackage{orcidlink}

\begin{document}
\title{Instantons, fluctuations and singularities in the supercritical stochastic  nonlinear Schr\"odinger equation}

\author{Sumeja Burekovi\'c\,\orcidlink{0009-0004-1600-2112}}
\email{sumeja.burekovic@ruhr-uni-bochum.de}
\affiliation{Institute for Theoretical Physics I, Ruhr University Bochum,
Bochum, Germany}
\author{Tobias Sch\"afer\,\orcidlink{0000-0002-9255-4718}}
\email{tobias.schaefer@csi.cuny.edu}
\affiliation{Department of Mathematics, College of Staten Island,   
Staten Island, NY, USA \& Physics Program, CUNY Graduate Center, 
New York, NY, USA}
\author{Rainer Grauer\,\orcidlink{0000-0003-0622-071X}}
\email{grauer@tp1.rub.de}
\affiliation{Institute for Theoretical Physics I, Ruhr University Bochum,
Bochum, Germany}

\date{\today}

\begin{abstract}
Recently, Josserand et al.~proposed a stochastic nonlinear Schr\"odinger model for finite-time singularity-mediated turbulence~[Phys.\,Rev.\,Fluids 5, 054607 (2020)]. Here, we use instanton calculus to quantify the effect of extreme fluctuations on the statistics of the energy dissipation rate. While the contribution of the instanton alone is insufficient, we obtain excellent agreement with direct simulations when including Gaussian fluctuations and the corresponding zero mode. Fluctuations are crucial to obtain the correct scaling when quasi-singular events govern the turbulence statistics.
\end{abstract}

\maketitle

\textit{Introduction} Understanding non-Gaussian statistics and anomalous scaling in turbulent systems is one of the outstanding challenges in classical physics~\cite{frisch-1995}. 
Given that the underlying probability distributions in such turbulent systems are dominated by extreme fluctuations, different toy models have been proposed, among them the stochastic  Burgers~\cite{burgers-1948}, the Kuramoto--Sivashinsky~\cite{manneville-1981}, and, more recently, the nonlinear Schr\"odinger equation (NLS)~\cite{chung-2009,josserand-2020}. In this letter, our work focuses on the NLS, although the presented methods can be applied to a wide range of turbulent models. Recall that the NLS naturally arises in a variety of contexts, in particular in nonlinear optics~\cite{zakharov-1972a, hasegawa-kodama:1995, agrawal-2019} and plasma physics~\cite{zakharov-1972b,novikov-1984}, and has come into the focus of novel applications in the context of Bose--Einstein condensates~\cite{chung-2009}, optical turbulence~\cite{dyachenko-newell-etal:1992}, and rogue waves~\cite{dematteis-2019a}. 

The one-dimensional stochastic focusing NLS on a spatial domain of size $\ell$, as introduced in~\cite{josserand-2020}, is given by
\begin{align} \label{eq:nls}
    &\partial_{t} \psi =\frac{\mathbbm{i}}{2} \partial_x^2 \psi + \mathbbm{i}\lvert\psi\rvert^{6} \psi-\nu \partial_x^4 \psi+  \chi^{1 / 2} * \eta\,,  \\
    &\psi(\cdot,t=0) = 0\,,  
\end{align}
where $*$ denotes spatial convolution and~$\chi^{1/2} * \chi^{1/2} = \chi$. 
The first three terms of equation~\eqref{eq:nls} describe the self-focusing conservative NLS with a supercritical nonlinearity~\cite{fibich-2015}. The remaining two terms constitute hyperviscous damping with hyperviscosity $\nu$ and a complex Gaussian forcing $\mathbb{E}\left[\eta(x,t) \eta^*(x^\prime,t^\prime)\right] =  2\sigma^2 \, \delta(x-x^\prime) \delta(t-t^\prime)$ that is white in time and has large-scale spatial correlations $\chi$ with amplitude $\sigma$. 

In~\cite{josserand-2020} it was proposed that the nearly singular collapsing solutions of the NLS~\eqref{eq:nls} provide a skeleton for the emergence of intermittency in the strongly turbulent case. 
In the present paper, we will analyze  the turbulence statistics of the NLS~\eqref{eq:nls} 
via instanton calculus to support the notion of singularity mediated turbulence as introduced in~\cite{josserand-2020}.  
\par 
The general picture of~\cite{josserand-2020} is consistent with the reasoning in hydrodynamic and magnetohydrodynamic turbulence, where the nearly singular structures such as shocks, vortices, or current sheets play a similar role. The important role of structures in understanding turbulent systems has already been suggested in the justification of the multifractal picture of turbulence~\cite{frisch-parisi:1985} and later incorporated in phenomenological models of turbulence~\cite{she-leveque:1994, grauer-krug-marliani:1994} and in the understanding of anomalous dissipation~\cite{duchon-robert:2000}. Another indication of the importance of the nearly singular structures is found in~\cite{zikanov-thess-grauer:1997}, where a combination of local and nonlocal nonlinearity allows the regularity of the singularity to change, leading to intermittency of varying strength. 
\par 
The NLS turbulence differs from the usual Navier--Stokes turbulence in two major ways. First, the NLS has two (instead of one, as in Navier--Stokes turbulence) independent dimensionless parameters:  
In equation~\eqref{eq:nls}, one can specify the system size and choose the viscosity~$\nu$ and forcing strength~$\sigma$ as independent parameters. Another possibility would be to fix  $\nu$ and vary the system size and~$\sigma$. This has also been discussed for the two-dimensional focusing NLS~\cite{amauger-josserand-etal:2023}.

The second, more important difference to the Navier-Stokes equations is the existence of a blow-up criterion for the critical mass in the focusing NLS, which determines whether initial configurations remain regular or form singularities in finite time (cf.~\cite{fibich-2015}). 
This particular property of the NLS has a significant impact on turbulence characteristics such as the probability distribution of local energy dissipation, as will become evident when discussing the main result.
\par 
This type of intermittency mediated by singularities seems to be an ideal candidate for the application of instanton calculus, since  instantons and their associated fluctuations capture the essence of singularity dominated turbulence. In contrast to fluid turbulence, where many other complex dynamical processes can occur in addition to the development of vortex tubes or sheets, e.g. topology-changing reconnection events, the tendency for singularity formation in the NLS is so robust that it dominates the turbulent dynamics. Our analysis supports this viewpoint and shows that NLS turbulence is accessible through the instanton approach, when including Gaussian  fluctuations and zero modes. 
\par 
In this letter, we quantify the effect of extreme fluctuations on NLS turbulence by considering extreme values of the energy dissipation density $\varepsilon$~\cite{josserand-2020}, defined as 
\begin{align}\label{eq:observable}
    \varepsilon(x,t) = 2 \nu \lvert\partial_x^2 \psi(x,t)\rvert^2
    \, . 
\end{align}
We analyze the probability density function (PDF)~$\rho$ of~$\varepsilon$, as further statistical quantities can be derived from it. 
To compute~$\rho$, we apply the instanton approach (cf.~\cite{grafke-2015c} and references therein), which consists of the following steps: i) find the instanton as the minimizer of the action in the corresponding path integral, ii) compute the Gaussian fluctuations around the instanton that lead to a fluctuation determinant, iii) add contributions from possible zero modes. This approach corresponds to large deviation theory  in mathematics~\cite{varadhan-1984, freidlin-2012}. While the road map i)--iii) has been known -- in principle -- for  decades, this program could not be successfully applied in the context of turbulent systems due to the lack of appropriate numerical methods and computational power. Nowadays, however, both are available.

All of these steps i)--iii) can be carried out in different ways, and each method has its advantages for a particular application. The path integral for stochastic differential equations can e.g.~be formulated as the Onsager--Machlup path integral~\cite{onsager-machlup:1953,machlup-onsager:1953} or the Janssen--de Dominicis path integral~\cite{janssen:1976,dedominicis:1976} by applying a Hubbard--Stratonovich transformation. 
In the following, limiting expressions for the PDF are derived analytically and evaluated numerically. 
Here, we use the path integral over all noise realizations, since the noise considered here mimics a large-scale forcing and hence, this formulation has computational advantages over the more general approaches~\cite{schorlepp-2021,grafke-schaefer-vanden-eijnden:2023,bouchet-reygner:2022}
when calculating the fluctuation determinant (see~\cite{schorlepp-2023b} for details). Thus, the 
PDF of the energy dissipation density $\varepsilon$ at a value $a$  reads:  
\begin{align} \label{eq:path integral}
    \rho(a)
    &=  \int \mathcal{D} \eta \, \exp\left(-\frac{1}{2\sigma^2}  \left\lVert\eta\right\rVert^2_{L^2}\right)
    \, \delta(F[\eta] - a)
    \,,
\end{align}
with $\langle\cdot,\cdot \rangle_{L^2}$ denoting the~$L^2$ product in space and time. 
The central object in this path integral is the solution map $F$ that solves the NLS~\eqref{eq:nls} for a given input noise $\eta$, and returns the observable value: $F[\eta] = O[\psi[\eta](\cdot,T)]$ at a final time $T>0$. Here, we take the energy dissipation density as $O[\psi(\cdot, T)]=\varepsilon(0,T)$. 
\\
In the constraint $O[\psi(\cdot,T)]= a$, the interest is in extreme events, i.e.~in large values of $a$ as it was introduced for Burgers turbulence~\cite{gurarie-1996, falkovich-1996, balkovsky-1997,chernykh-2001} almost 30 years ago.
Formally, we take the small noise limit  $\sigma \downarrow 0$, which is equivalent as long as $a$ is sufficiently large~(cf.~\cite{schorlepp-2021}). 
By Laplace's method, the path integral~\eqref{eq:path integral} will have the following asymptotic form:
\begin{align}\label{eq:instanton pdf subexponential prefactor}
    \rho(a)  = C(a) \exp\left(-\frac{S_{\mathrm{I}}(a)}{\sigma^2}\right) 
    \left(1 + \mathcal{O}\left(\sigma^2\right)\right) 
    \,,
\end{align}
as $\sigma \downarrow 0$, where we call $C(a)$ the algebraic prefactor, and the exponential contribution stems from the instanton. 
While there are results on instantons in the NLS~\cite{falkovich-2001, terekhov-2014,  onorato-2016, poppe-2018,dematteis-2019a, roberti-2019, alqahtani-2021} neither the specific supercritical form~\eqref{eq:nls}, nor the energy dissipation density observable, have been considered so far.  
More importantly, to the best of our knowledge, the computation of the PDF prefactor $C$ for the NLS~\eqref{eq:nls} is a novelty as well and turns out to be crucial. In the following, we briefly explain how we obtain both the instanton and the prefactor contributions.  
\par 
\textit{i) Instanton}  
In the path integral~\eqref{eq:path integral}, the instanton is the minimizer of the action functional with observable constraint: 
\begin{equation} \label{eq:instanton action minimum}
    S_{\mathrm{I}}(a) = \min_{\eta \text{ s.t.\,} F[\eta]= a} S[\eta] , \quad S[\eta] = \frac{1}{2} \left\lVert\eta\right\rVert^2_{L^2}
    \, .
\end{equation}
For increased numerical efficiency and stability, in order to find the optimal $\eta$, we do not directly solve the corresponding Euler--Lagrange or instanton equations, but use optimal control methods with control variable~$\eta$, similarly to~\cite{schorlepp-2022}. 
By this we iteratively solve a deterministic forward and backward PDE of very similar shape and perform unconstrained optimization as follows:      
\par 
First, we write the constraint in equation~\eqref{eq:instanton action minimum} as a Lagrange term in the functional $R_\lambda[\eta] = S[\eta] - \lambda (F[\eta] - a)$ with a Lagrange multiplier $\lambda$. Since the instanton action~$S_{\mathrm{I}}$ will turn out to be non-convex in~$a$ (cf.~\cite{alqahtani-2021}), we use an augmented Lagrangian~\cite{nocedal-2006}  with a penalty parameter~$\mu$:  
\begin{align}
    R[\eta]  = S[\eta]
    - \lambda (F[\eta] - a)
    + \frac{\mu}{2} \left(F[\eta] - a\right)^2 
    \,. 
    \label{eq:target functional}
\end{align}
For sufficiently large $\mu$, the instanton solution of equation~\eqref{eq:instanton action minimum} for an observable value~$a$, is $\eta_a = \mathrm{argmin}_\eta R[\eta]$.   
We perform a gradient-based minimization of~$R$, thus we need $\delta F/ \delta\eta$. 
To evaluate this complicated expression, we employ the adjoint-state method~\cite{plessix-2006} by using a field-valued Lagrange multiplier~$z$. Details are given in appendix~A. 
In total, the gradient reads: 
\begin{align}\label{eq:gradient}
    \frac{\delta R}{\delta \eta}  = \eta - \chi^{1/2} * z\,,     
\end{align}
where~$z$ solves: 
\begin{align}\label{eq:adjoint pde}
    \partial_t z -  \frac{\mathbbm{i}}{2} \partial_x^2 z - \nu \partial_x^4 z - 4 \mathbbm{i}  \left\lvert\psi\right\rvert^{6} z + 3 \mathbbm{i}  \left\lvert\psi\right\rvert^{4} \psi^2 z^*  = 0\,, 
    \\[0.4em]
    z(x, T)  =  4 \nu \big(
    \lambda 
    - \mu (\varepsilon(0,T) -a )\big) \, 
    \partial_x^2 \psi (0,T) \, \delta''(x)
   \, ,
\end{align}
with $\psi = \psi[\eta]$ through equation~\eqref{eq:nls}. 
At the instanton, by the first-order optimality condition, the gradient~\eqref{eq:gradient} is zero, i.e.~$\eta_a = \chi^{1/2}*z_a$. Substituting this expression for~$z$ in the adjoint PDE~\eqref{eq:adjoint pde} yields the instanton/Hamilton equations  with  the optimal~$\eta_a$ corresponding to the optimal conjugated momentum of the system up to a factor~$\chi^{1/2}$.  
\par 
For a fixed observable value $a$, the instanton $\eta_a$, from which $S_{\mathrm{I}}(a)$ is obtained,  is found by 
solving a series of unconstrained optimization problems $\min_\eta R[\eta]$ for increasing values $\mu^{(i)}$ of the penalty parameter, and the Lagrange multiplier is updated according to~\cite[p.~515]{nocedal-2006}. 
For each evaluation of the gradient~\eqref{eq:gradient}, we first solve the forward equation~\eqref{eq:nls} and then the adjoint equation~\eqref{eq:adjoint pde}. 
\par 
We implemented a pseudospectral code to solve these equations with a $1/4$ anti-aliasing~\cite{fructus-2005}.  In line with~\cite{schorlepp-2022}, we use the L-BFGS scheme~\cite{liu-1989} for the minimization in real variables. 
For this, we write all complex fields as two-dimensional real vectors, as shown in appendix~C. 
The optimization code has been consistently discretized, with the Heun scheme with integrating factor for the forward equation.   
\\ 
\textit{ii)--iii) Gaussian fluctuations and zero mode} 
Now, we compute an estimate of the prefactor $C$ in equation~\eqref{eq:instanton pdf subexponential prefactor}. All formulae are given for $a>0$. For this computation, we employ the approach based on Fredholm determinants 
established in~\cite{schorlepp-2023b}. After calculating the instanton solutions $\psi_a^\varphi$, $\eta_a^\varphi$ and $z_a^\varphi$, we insert these fields as background fields in the computation of the prefactor. 
Since the NLS~\eqref{eq:nls} as well as the observable function~\eqref{eq:observable} are globally $U(1)$ invariant with respect to the complex phase $\varphi = \mathrm{arg}(\psi)$, the instanton solution is degenerate in~$\varphi$ and therefore gives rise to a zero mode which we indicate by the superscript~$\varphi$. 
Due to the zero mode, the Fredholm determinant in~\cite{schorlepp-2023b} is ill-defined and has to be regularized. For this, we follow~\cite{schorlepp-2023a}.  
\par 
We split the domain of integration of the path integral~\eqref{eq:path integral} into the submanifold~$M^1$ of the instanton (noise) solutions~$\eta_a^\varphi$, and the subspace~$N_\varphi M^1$ that is normal (with respect to the~$L^2$ product) to the zero mode: $\eta \to \eta_a^\varphi + \sigma \Tilde{\eta}$.  
The submanifold $M^1 = \mathrm{argmin}_\eta S[\eta]$ is one-dimensional since the zero mode stems from the scalar parameter~$\varphi$. 
The split of integration directions is usually done formally using the Faddeev--Popov method~\cite{faddeev-1967}. 
As detailed in appendix~B, the leading-order  prefactor~$C$ in formula~\eqref{eq:instanton pdf subexponential prefactor} then reads: 
\begin{align}
     C(a) = 
    \frac{1}{\sigma^2}  \, \lvert\lambda_a\rvert \, \mathrm{det}'(\mathrm{Id} - B_a)^{-1/2} 
    \,. 
    \label{eq:prefactor formula}
\end{align}
The Lagrange multiplier $\lambda_a = \mathrm{d} S_{\mathrm{I}}/ \mathrm{d} a$ 
is obtained from the optimization scheme given above.  
The regularized Fredholm determinant~$\mathrm{det}'$ is approximated using the largest eigenvalues~$\kappa_a^{(i)} \neq 1$ of  the  operator~$B_a$, given by 
\begin{align}
    B_a =\left.\lambda_a \operatorname{pr}_{\eta_a^{\perp}} \frac{\delta^2 F}{\delta \eta^2}\right\rvert_{\eta = \eta_a} \operatorname{pr}_{\eta_a^{\perp}}
    \,,  
    \label{eq:definition integral operator fredholm determinant}
\end{align}
with the projection operator~$\operatorname{pr}_{\eta_a^{\perp}}$ defined in appendix~B. 
These eigenvalues are calculated iteratively~\cite{lehoucq-1998}
from the solution of  second-order equations~\cite{schorlepp-2023b}, which are given in appendix~C.  In figure~\ref{fig:fredholm det}, we show the convergence of the numerical approximation of 
$\mathrm{det}'(\mathrm{Id} - B_a)$ for an example observable value. 
\begin{figure}
\includegraphics[width=0.48\textwidth]{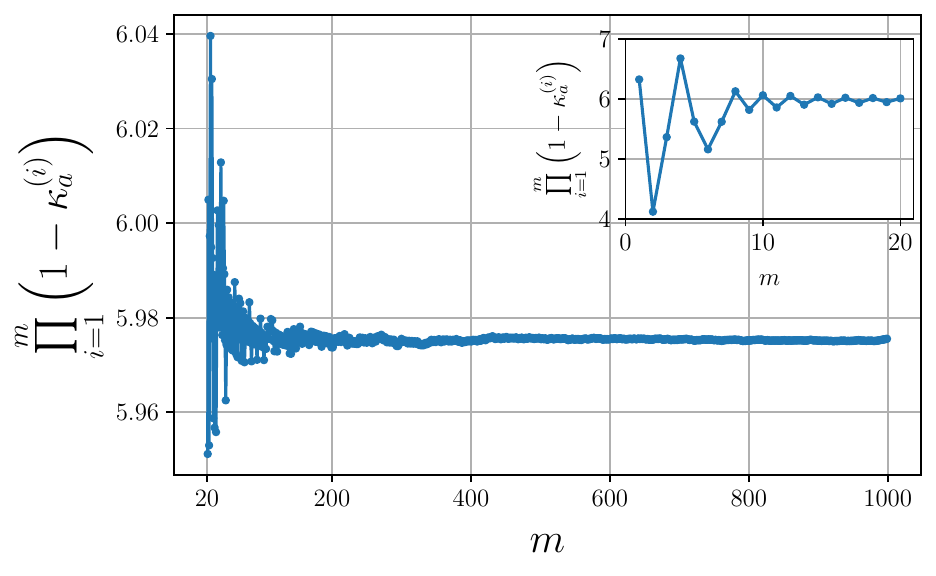}
    \caption{
    Result of numerically computing the $m = 1000$ eigenvalues~$\kappa_a^{(i)}$ of $B_a$ with the largest absolute value for $\varepsilon(0,T) = a = 0.015$. 
    The figure shows the finite product \scalebox{0.75}{$\prod\limits_{i=1}^m \left(1- \kappa_a^{(i)}\right)$} approximating the regularized Fredholm determinant $\det'(\mathrm{Id} - B_a)$  
    without the zero mode $\kappa_a = 1$. The numerical approximation of $\det'(\mathrm{Id} - B_a)$ converges quickly. 
    All subsequent results in this letter are obtained for $m=1000$. 
}
\label{fig:fredholm det}
\end{figure}
\par 
\textit{Direct Numerical Simulations (DNS)}
We also performed Monte--Carlo simulations of the NLS~\eqref{eq:nls} with the same parameters as for the instanton and fluctuation computations, which are given in table~\ref{tab:parameters}. Following~\cite{josserand-2020}, we only force large length scales:  $ \widehat{\chi}_k  = \frac{1}{\ell}  \mathbbm{1}_{\left( 0 < \lvert k \rvert  < 0.3 \frac{\ell}{2 \pi}\right)}$ for $k \in \mathbb{Z}$. 
\begin{table}[t]
    \renewcommand{\arraystretch}{1.2} 
    \centering
    \caption{Parameters that enter the DNS, the optimization scheme for the instanton computation, and the computation of the Gaussian fluctuations.}
    \begin{ruledtabular}
        \begin{tabular}{cp{5cm}c}
            \textbf{Parameter} & \textbf{Definition} & \textbf{Value} \\
            \hline
            $\ell$ & Length of periodic spatial domain   & $153.6$ 
            \\
            $\nu$ & Hyperviscosity & $10^{-2}$
            \\
            $T$ & Time interval $[0,T]$ & $2.0$
            \\
            $N_t$ & Time resolution & $2^{12}$ 
            \\
            $N_x$ & Spatial resolution  &$2^{12}$
            \\ 
        \end{tabular}
    \end{ruledtabular}
    \label{tab:parameters}
\end{table}
To compute the PDF~$\rho$ from DNS data, we used $9.3 \cdot 10^6$ simulated fields and evaluated $3.7 \cdot 10^8$ statistically independent samples. 
\par
\begin{figure}
    \includegraphics[width=0.48\textwidth]{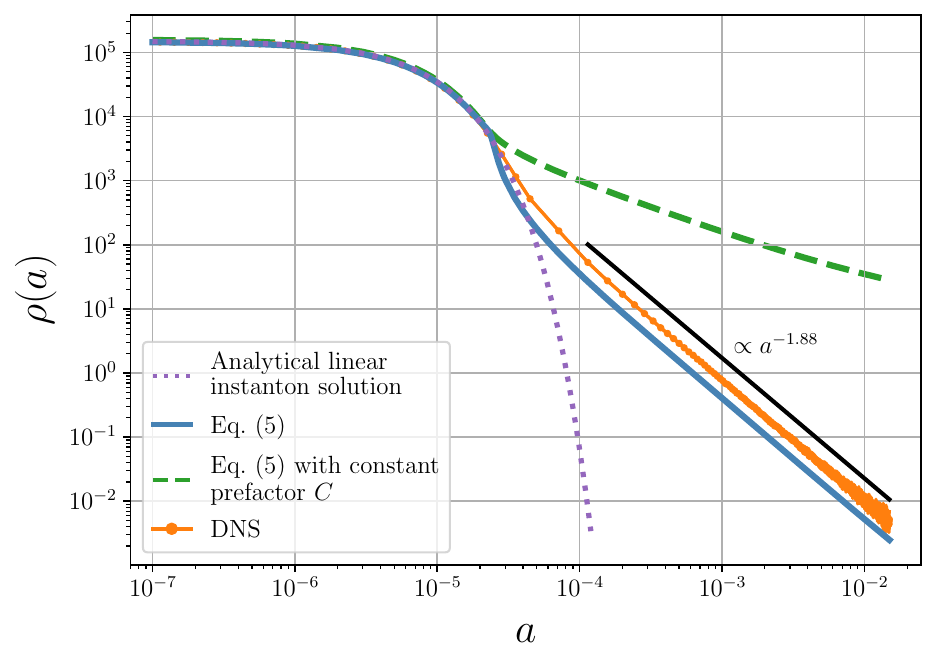}
    \caption{Comparison of instanton predictions and DNS results for the PDF~$\rho$ of the energy dissipation density~\eqref{eq:observable}.  The dashed line indicates the leading order contribution $\exp(-S_{\mathrm{I}}/\sigma^2)$ in equation~\eqref{eq:instanton pdf subexponential prefactor} with a constant prefactor instead of equation~\eqref{eq:prefactor formula}.  
    The shaded regions for the DNS data are $99\,\%$ Wilson score intervals~\cite{brown-2001}. The DNS data and full instanton prediction show a power-law decay for large~$a$ with scaling exponent $\approx 1.88$.}
\label{fig:pdfs}
\end{figure}

\begin{figure}
    \includegraphics[width=0.45\columnwidth]{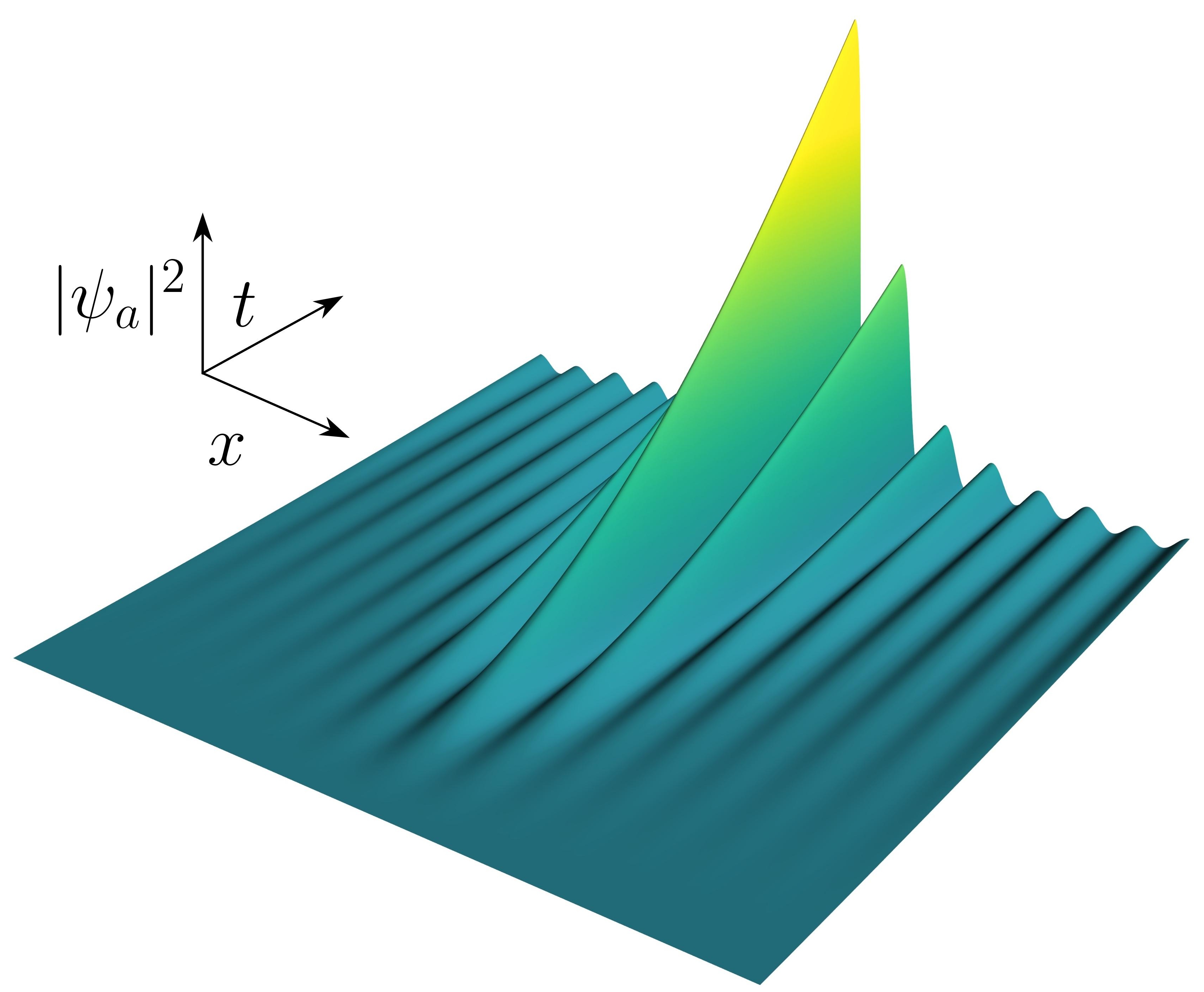}
    \includegraphics[width=0.45\columnwidth]{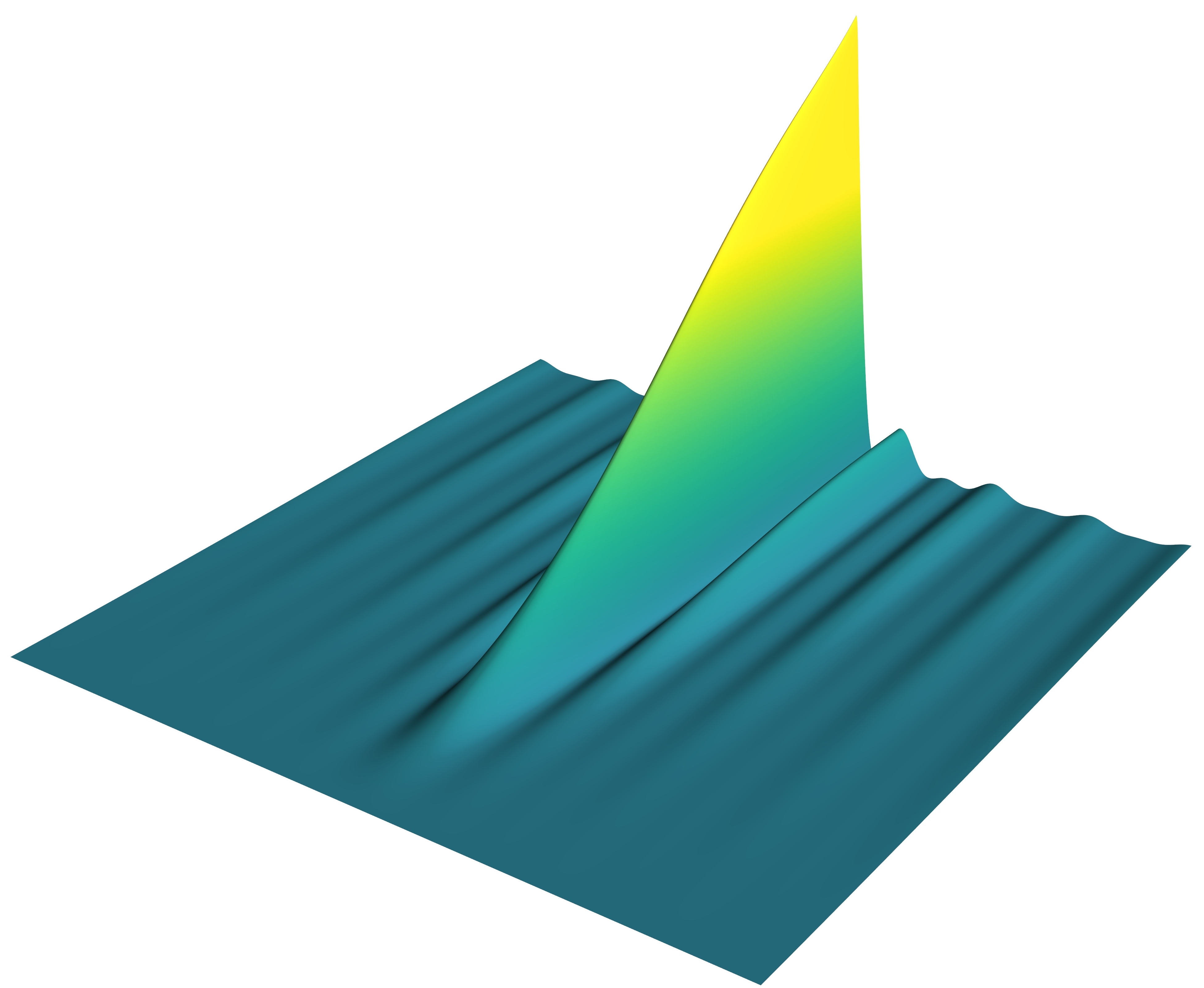} 
    \caption{Spatio-temporal surface plots of instanton fields $\psi_a$ for $a=2\cdot 10^{-5}$ (left) and $a=0.015$ (right). As a scale reference, $\mathrm{max}_{x,t}\lvert\psi_a\rvert^2 \approx 0.34$ (left) and $\mathrm{max}_{x,t}\lvert\psi_a\rvert^2 \approx 1.55$ (right).}
    \label{fig:instantons}
\end{figure}

 
\textit{Discussion of results}
The main result is shown in figure~\ref{fig:pdfs}, which displays the PDF for the energy dissipation density~\eqref{eq:observable}. In this figure, the noise strength is given by $\sigma^2=0.5$. Other values of the noise strength ($\sigma^2=0.4, \sigma^2=0.75$) were studied as well and show a very similar behavior.
Two regions can be identified: a region belonging to smaller values of energy dissipation ($a \le 2\cdot 10^{-5}$) and a region of rare fluctuations of the energy dissipation ($4\cdot 10^{-5} \le a  \le 10^{-2}$).
In the first region, the PDF calculated from the DNS agrees almost exactly with the asymptotic prediction of the instanton calculation including the fluctuations. However,  even the prediction of the PDF using solely the instanton with a constant prefactor~$C$ in equation~\eqref{eq:instanton pdf subexponential prefactor}, instead of the full expression~\eqref{eq:prefactor formula}, yields a result indistinguishable from the DNS. A typical instanton evolution is shown in the left part of figure~\ref{fig:instantons}.
\par 
To understand this observation, we analytically solved the instanton equations and computed the prefactor in case of vanishing nonlinearity, which is given in appendix~D. The resulting prediction for the PDF, an exponential distribution, is also shown in figure~\ref{fig:pdfs} and again agrees  well with the DNS result in the first region. 
This can be explained by the fact that in the first region, the corresponding instanton is below the critical mass that would lead to a collapse, such that the focusing nonlinearity effectively vanishes. We tested this hypothesis numerically by setting the instantons in this parameter range as initial conditions in the deterministic conservative NLS ($\sigma = \nu = 0$). Indeed, the numerical solution produced no collapse in this region. This range therefore corresponds to an almost linear regime in which the instanton prediction is nearly exact, i.e.~the small noise limit is almost perfectly fulfilled. 
\par 
The result in the second region is even more surprising. In this region, there is initially no agreement between the results of the DNS and the prediction of the PDF by the instanton without the fluctuations. Note that this curve can be shifted arbitrarily on the vertical axis, since the normalization of the PDF is not determined by the instanton itself. 
In this range ($4\cdot 10^{-5} \le a \le 10^{-2}$), the PDF of the energy dissipation shows a power law  behavior~(cf.~\cite{chung-2009}), i.e.~it is completely dominated by the prefactor and the exponential part due to the instanton alone is subdominant. The agreement of the power law scaling prediction of the instanton calculation including the fluctuations and the DNS is almost perfect (with convergence of the two curves –– instanton prediction and  extrapolated DNS data –– for even larger~$a$). 
Our interpretation of this result is given by the special characteristic of the focusing NLS to form strongly localized structures. Also, in this parameter range, we used the instantons as initial conditions in the conservative and deterministic NLS. Here, unlike the first region, these initial conditions led to a collapse. A typical instanton in this regime is depicted in the right of figure~\ref{fig:instantons}. In contrast to other turbulent systems, the NLS turbulence is characterized by the occurrence of localized, spatially barely interacting nearly singular structures. This is particularly well illustrated in figure~1 of the work of Josserand et al.~\cite{josserand-2020}. We can also interpret this property of the collapsing NLS as the fact that the action landscape in the path integral formulation exhibits strongly localized extrema, which can be very well represented by a Gaussian approximation around the instanton.
\par 
\textit{Conclusions and Outlook}
The main result of this letter is that the instanton formalism captures the statistics of the supercritical stochastic NLS. 
The formalism can precisely describe the PDF of the energy dissipation~\eqref{eq:observable}, which is far from a Gaussian distribution, and gives the correct scaling in the strongly nonlinear region. The decisive factor was the inclusion of the Gaussian fluctuations around the instanton and the consideration of the zero mode in our analysis. 
In some way, this case can be considered as a paradigm for turbulence that is dominated by isolated weakly singular structures. 
This differs from the situation in real turbulence as it occurs, for example, in the Navier--Stokes equations. But even in the simpler case of shock-dominated Burgers turbulence, the situation is more complex, so that there are significant deviations in the gradient statistics at higher Reynolds numbers between the predictions of the instanton formalism and the results of numerical simulations. For instance, shocks can merge in the Burgers turbulence and thus have a further influence on the gradient statistics. The situation is even more complex in Navier--Stokes turbulence, where reconnection of vortex tubes and other, more complex processes can occur. Such events are not yet included in the instanton formalism. 
The link between Gaussian PDFs in the realm of small observable values and the PDFs associated with large observable values dominated by singularities remains elusive. 
To move in this direction, one possibility would be to weaken the nonlinearity in the NLS towards the critical case. This would reduce the nature of the extreme singularities and could therefore systematically lead towards more complex interactions. Research in this direction is in progress. 
\par 
\textit{Acknowledgment}
The authors would like to thank  Timo Schorlepp for helpful discussions on the Fredholm determinant prefactor. 
R.G.~acknowledges support from the German Research Foundation DFG within the Collaborative Research Center SFB1491. T.S.~acknowledges support from the NSF grant DMS-2012548. 

\bibliographystyle{apsrev4-2}
\bibliography{ref.bib}

\onecolumngrid
\newpage 
\appendix 
\setcounter{equation}{0}
\renewcommand{\theequation}{A.\arabic{equation}}
\section{Appendix A: Application of the adjoint-state method to the NLS}
To compute the gradient $\delta R/ \delta\eta$ in equation~\eqref{eq:target functional}, we employ the adjoint-state method~\cite{plessix-2006}.
In the following, we  give the concrete application of the method to the NLS system. 
We add the state equation, in our case the NLS~\eqref{eq:nls},  by a field-valued Lagrange multiplier~$z$ to the original target functional and consider the fields~$\psi$ and~$\eta$ to be independent from each other, temporarily enlarging the state space: 
\begin{equation}
\begin{aligned}
     \mathcal{L}[\psi, \eta, z] &= \frac{1}{2} \left\lVert\eta\right\rVert^2_{L^2} - \lambda (O[\psi(\cdot,T)] - a)
    + \frac{\mu}{2} \left(O[\psi(\cdot,T)] - a\right)^2  \\
    &+ \left\langle z, \partial_{t} \psi-\frac{\mathbbm{i}}{2} \partial_x^2 \psi - \mathbbm{i}\lvert\psi\rvert^{6} \psi+\nu \partial_x^4 \psi- \chi^{1 / 2} * \eta\right\rangle_{L^2} \, .
\end{aligned}
\end{equation}
Observe that the target functional~$R$ is recovered by setting $\psi = \psi[\eta]$, i.e.~$\psi$ depending on~$\eta$ through the state equation.  
The gradient in~$\eta$ for any~$z$ becomes: 
\begin{align}
        \frac{\delta R[\eta]}{\delta \eta} 
        =  \frac{\delta \mathcal{L}[\psi[\eta],\eta,z]}{\delta \eta}
        = \left.\frac{\delta \mathcal{L}[\psi,\eta,z]}{\delta \psi}\right\rvert_{\psi = \psi[\eta]} \, \frac{\delta \psi}{\delta \eta} 
        +  \left.\frac{\delta \mathcal{L}[\psi,\eta,z]}{\delta \eta}\right\rvert_{\psi = \psi[\eta]}
        \, .
        \label{eq:gradient of target functional}
\end{align}
To avoid having to compute the Jacobian $\delta \psi/ \delta \eta$ in equation~\eqref{eq:gradient of target functional}, the adjoint field~$z$ is now fixed by setting $\delta \mathcal{L}/\delta \psi = 0$. 
This gives the adjoint equation whose natural direction is backward in time: 
\begin{align}\label{eq:adjoint equation}
    &\partial_t z -  \frac{\mathbbm{i}}{2} \partial_x^2 z - \nu \partial_x^4 z - 4 \mathbbm{i}  \left\lvert\psi\right\rvert^{6} z + 3 \mathbbm{i}  \left\lvert\psi\right\rvert^{4} \psi^2 z^*  = 0\,, 
    \\
    &z(\cdot, T)  =  \big(
    \lambda 
    - \mu (O[\psi[\eta] (\cdot,T)] -a )\big) \, 
    \left. \frac{\delta O}{\delta \psi}\right\rvert_{\psi = \psi[\eta] (\cdot,T)} \,, 
\end{align}
where, here, $O[\psi(\cdot, T)]=\varepsilon(0,T)$ for the energy dissipation density~\eqref{eq:observable},  and hence:
\begin{align}
    \left. \frac{\delta O}{\delta \psi}\right\rvert_{\psi = \psi[\eta] (\cdot,T)} = 4 \nu \partial_x^2 \psi (0,T) \,  \delta''(x) \,. 
\end{align}
The gradient becomes: 
\begin{align}\label{eq:gradient optimization}
\frac{\delta R}{\delta \eta} = \eta -
 \big(
    \lambda 
    - \mu (O[\psi[\eta] (\cdot,T)] -a )\big) \frac{\delta F}{\delta \eta} 
 \overset{\eqref{eq:gradient of target functional}}{=} \frac{\delta \mathcal{L}}{\delta \eta}
 = \eta - \chi^{1/2} * z    \,, 
\end{align}
which gives equation~\eqref{eq:gradient} in the main text. 
\setcounter{equation}{0}
\renewcommand{\theequation}{B.\arabic{equation}}
\section{Appendix B: Derivation of the PDF with Gaussian fluctuations and zero mode}
In this section, we sketch how to obtain the PDF~$\rho(a) = \mathbb{E}\left[\delta(F[\eta] -a) \right]$ of the energy dissipation density~\eqref{eq:observable} from the path integral~\eqref{eq:path integral} by including Gaussian fluctuations around the instanton and the zero mode. First, we use the Fourier representation of the delta function. 
Then, we perform a change of variables in the path integral~\eqref{eq:pdf change of variables 1}: we split the domain of integration into the submanifold~$M^1$ of the instanton (noise) solutions~$\eta_a^\varphi$ which minimize~$R_\lambda$, and the subspace~$N_\varphi M^1$ that is normal (with respect to the~$L^2$ product) to the zero mode: $\eta \to \eta_a^\varphi + \sigma \Tilde{\eta}$. In the limit $\sigma \to 0$, the original domain of integration is retained under this  transformation: 
\begin{align}
  \rho(a)
    &= \frac{1}{2 \pi \mathbbm{i}\sigma^2 }  \int_{-\mathbbm{i} \infty}^{\mathbbm{i} \infty} \mathrm{d} \lambda \,  \mathbb{E}\left[\exp\left(\frac{\lambda}{\sigma^2}(F[\eta] -a)\right)  \right]
    = \frac{1}{2 \pi \mathbbm{i}\sigma^2 }  \int_{-\mathbbm{i} \infty}^{\mathbbm{i} \infty} \mathrm{d} \lambda 
    \int \mathcal{D} \eta 
    \exp\left(-\frac{1}{\sigma^2}R_{\lambda}[\eta]\right)  
    \label{eq:pdf change of variables 1}
    \\[0.4em] 
    &= \frac{1}{2 \pi \mathbbm{i}\sigma^2 } \frac{1}{(2 \pi \sigma^2)^{1/2}}
    \int_{-\mathbbm{i} \infty}^{\mathbbm{i} \infty} \mathrm{d} \lambda \int_{M^1} \mathcal{D} \eta_a^\varphi \int_{N_\varphi M^1} \mathcal{D} \Tilde{\eta} \,
    \exp\left(-\frac{1}{\sigma^2} R_{\lambda} \left[\eta_a^\varphi + \sigma \Tilde{\eta}\right] \right)
    \left(1 + \mathcal{O}\left(\sigma^2\right)\right)  
     \,, 
\end{align}
where the normalization factor $(2\pi \sigma^2)^{1/2}$ stems from the fact that only the normal direction is scaled by~$\sigma$, and the action functional reads
\begin{align}
    R_{\lambda}[\eta] = \frac{1}{2} 
     \left\lVert\eta \right\rVert^2_{L^2} - \lambda \left(F\left[\eta\right] - a\right)
     \,, 
\end{align}
with the Lagrange multiplier~$\lambda$, as in the main text. 
After changing variables, $\lambda \to \lambda_a + \sigma \Tilde{\lambda}, \eta_a^\varphi \to \varphi$, the PDF becomes:
\begin{align}
     \rho(a)
     &= \frac{1}{2 \pi \mathbbm{i}\sigma^2}  \frac{1}{(2 \pi)^{1/2}} \int_0^{2 \pi} \mathrm{d} \varphi \, 
     \left\lVert\frac{\partial \eta_a^\varphi}{\partial \varphi}\right\rVert_{L^2}
     \int_{-\mathbbm{i} \infty}^{\mathbbm{i} \infty} \mathrm{d} \Tilde{\lambda}  
     \int_{N_\varphi M^1} \mathcal{D} \Tilde{\eta} \, 
     \exp\left(-\frac{1}{\sigma^2} R_{\lambda_a + \sigma  \tilde{\lambda}}\left[\eta_a^\varphi + \sigma \Tilde{\eta}\right]\right)
     \left(1 + \mathcal{O}\left(\sigma^2\right)\right)
     \,. 
     \label{eq:pdf change of variables 2}
\end{align}
Now, we apply Laplace's method: The action functional~$R_{\lambda_a + \sigma  \tilde{\lambda}}\left[\eta_a^\varphi + \sigma \Tilde{\eta}\right]$ is expanded up to second order around the stationary point/instanton $(\lambda_a, \eta_a^\varphi)$, i.e.~for $\sigma \downarrow 0$: The first-order terms vanish due to the optimality condition, hence, 
with $R_{\lambda_a}[\eta_a^\varphi] = S_{\mathrm{I}}(a)$: 
\begin{align}
    & R_{\lambda_a + \sigma \tilde{\lambda}}\left[\eta_a^\varphi + \sigma \Tilde{\eta}\right]
    = S_{\mathrm{I}}(a)  
    +  \frac{\sigma^2}{2}  
    \left\langle\Tilde{\eta},
    \left[\mathrm{Id} - \lambda_a  \left.\frac{\delta^2 F}{\delta \eta^2}\right\rvert_{\eta = \eta_a^\varphi} \right]
    \Tilde{\eta} \right\rangle_{L^2} - \sigma^2 \Tilde{\lambda} \left\langle\Tilde{\eta},     
      \left.\frac{\delta F}{\delta \eta}\right\rvert_{\eta = \eta_a^\varphi}  \right\rangle_{L^2} 
      + \mathcal{O}\left(\sigma^3\right)
      \,. 
\end{align}
Substituting this expansion into equation~\eqref{eq:pdf change of variables 2} and using again the Fourier representation of the delta function, we find:
\begin{equation}
\begin{aligned}
    \rho(a) &= 
     \frac{1}{(2 \pi)^{1/2}} \frac{1}{\sigma^2} \int_0^{2 \pi} \mathrm{d} \varphi \, 
     \left\lVert\frac{\partial \eta_a^\varphi}{\partial \varphi}\right\rVert_{L^2}
     \exp\left(-\frac{S_{\mathrm{I}}(a)}{\sigma^2}\right) \frac{\lvert\lambda_a\rvert}{\left\lVert\eta_a^\varphi\right\rVert_{L^2}}
     \quad \times 
     \\[0.6em]
     & 
    \times \int_{N_\varphi M^1} \mathcal{D} \Tilde{\eta}
    \; \delta\left(\langle e_a^\varphi, \Tilde{\eta} \rangle_{L^2}\right)
    \exp\left(-\frac{1}{2} \left\langle\Tilde{\eta}, \left[\mathrm{Id} - \lambda_a \left.\frac{\delta^2 F}{\delta \eta^2}\right\rvert_{\eta = \eta_a^\varphi} \right] \Tilde{\eta} \right\rangle_{L^2}\right)
    \, 
    \left(1 + \mathcal{O}\left(\sigma^2\right)\right)
    \,.
    \label{eq:pdf gaussian integral}
\end{aligned}
\end{equation}
with  $e_a^\varphi = \eta_a^\varphi/ \lVert\eta_a^\varphi\rVert_{L^2}$. 
The last integral is a Gaussian integral in $\Tilde{\eta}$, where we exactly removed the direction which corresponds to the zero mode: From the instanton equation, we have: 
\begin{align}
    \left.\frac{\delta R_{\lambda_a + \sigma \tilde{\lambda}}\left[\eta_a^\varphi + \sigma \Tilde{\eta}\right]}{\delta \eta}\right\rvert_{\sigma=0}
    = \eta_a^\varphi - \lambda_a \left.\frac{\delta F}{\delta \eta}\right\rvert_{\eta = \eta_a^\varphi}
    = 0 \,.
\end{align}
Upon differentiating with respect to $\varphi$, we obtain: 
\begin{align}
    \left(\mathrm{Id}  - \lambda_a \left.\frac{\delta^2 F}{\delta \eta^2}\right\rvert_{\eta = \eta_a^\varphi}\right) \left(\frac{\partial \eta_a^\varphi}{\partial \varphi}\right) = 0
    \,. 
\end{align}
Thus, $\partial \eta_a^\varphi/\partial \varphi$ exactly corresponds to the zero mode of the operator
\begin{align}
    \left(\mathrm{Id}  - \lambda_a \left.\frac{\delta^2 F}{\delta \eta^2}\right\rvert_{\eta = \eta_a^\varphi}\right)
    \,, 
\end{align}
whose determinant is needed for the prefactor evaluation. 
But since the domain of integration is $N_\varphi M^1$, the tangent vector of $M^1$, $\partial \eta_a^\varphi/\partial \varphi$, is outside the domain of integration and the Gaussian integral in equation~\eqref{eq:pdf gaussian integral}  can now be calculated: 
\begin{align}
    &\int_{N_\varphi M^1} \mathcal{D} \Tilde{\eta} \; \delta(\langle e_a^\varphi, \Tilde{\eta} \rangle_{L^2}) \, 
    \exp\left(-\frac{1}{2} \left\langle\Tilde{\eta}, \left[\mathrm{Id} - \lambda_a \left.\frac{\delta^2 F}{\delta \eta^2}\right\rvert_{\eta = \eta_a^\varphi} \right] \Tilde{\eta} \right\rangle_{L^2}\right)
    = \frac{1}{(2 \pi)^{1/2}} \, \mathrm{det}'(\mathrm{Id} - B_a)^{-1/2}
    \,. 
    \label{eq:gaussian integral fredholm determinant int result 2}
\end{align}
The factor ${(2 \pi)^{-1/2}}$ stems from the fact that due to the delta function $\delta(\langle e_a^\varphi, \Tilde{\eta} \rangle_{L^2})$, one additional direction is removed from the integration, and thus, one degree of freedom is missing from the normalization. This results again in an additional normalization factor.    
In equation~\eqref{eq:gaussian integral fredholm determinant int result 2}, $\det'$ denotes the Fredholm determinant that is regularized by removal of its zero mode, indicated by the prime.   The operator $B_a$ is defined as
\begin{align}
    B_a =\left.\lambda_a \operatorname{pr}_{\eta_a^{\perp}} \frac{\delta^2 F}{\delta \eta^2}\right\rvert_{\eta = \eta_a^\varphi} \operatorname{pr}_{\eta_a^{\perp}}
    \,,  
    \label{eq:definition integral operator fredholm determinant appendix}
\end{align}
with the projection operator $\mathrm{pr}_{\eta_a^{\perp}}$ acting on an input $\delta \eta$ as: 
\begin{align}
    \left(\operatorname{pr}_{\eta_a^{\perp}} \delta \eta\right)(x, t)=\delta \eta(x, t)-\frac{\left\langle\eta_a^\varphi, \delta \eta\right\rangle_{L^2}}{\left\lVert\eta_a^\varphi\right\rVert_{L^2}^2} \, \eta_a^\varphi(x, t)
    \,, 
    \label{eq:definition projection operator fredholm determinant}
\end{align}
such that the condition $\tilde{\eta} \perp \eta_a^\varphi$ from the delta function in equation~\eqref{eq:gaussian integral fredholm determinant int result 2} is incorporated, and we obtain:
\begin{align}
    \rho(a) = 
    \frac{1}{2 \pi \sigma^2} \int_0^{2 \pi} \mathrm{d} \varphi \, 
    \left\lVert\frac{\partial \eta_a^\varphi}{\partial \varphi}\right\rVert_{L^2} \; 
    \exp\left(-\frac{S_{\mathrm{I}}(a)}{\sigma^2}\right) 
    \frac{\lvert\lambda_a\rvert}{\left\lVert\eta_a^\varphi\right\rVert_{L^2}} \, \mathrm{det}'(\mathrm{Id} - B_a)^{-1/2}
    \left(1 + \mathcal{O}\left(\sigma^2\right)\right)  
    \,. 
    \label{eq:pdf zero mode volume factor integral}
\end{align}
Thus, the zero mode introduces an additional volume factor which is now calculated. Due to the phase invariance of the observable function and the NLS, the instanton solution is phase-invariant and reads $\eta_a^\varphi = \eta_a^{\mathrm{ref}} \exp(\mathbbm{i} \varphi)$ and the PDF is asymptotically given by: 
\begin{align}
     \rho(a) &= 
    \frac{1}{\sigma^2}  \lvert\lambda_a\rvert \, \mathrm{det}'(\mathrm{Id} - B_a)^{-1/2} \exp\left(-\frac{S_{\mathrm{I}}(a)}{\sigma^2}\right) \left(1 + \mathcal{O}\left(\sigma^2\right)\right)
    \,. 
\end{align}
The leading-order  prefactor~$C$ in equation~\eqref{eq:instanton pdf subexponential prefactor} then reads: 
\begin{align}
     C(a) = 
    \frac{1}{\sigma^2}  \, \lvert\lambda_a\rvert \, \mathrm{det}'(\mathrm{Id} - B_a)^{-1/2} 
    \,,
\end{align}
which is equation~\eqref{eq:prefactor formula} in the main text. 
\setcounter{equation}{0}
\renewcommand{\theequation}{C.\arabic{equation}}
\section{Appendix C: Evaluation of the first and second variation}
The equations for the numerical optimization scheme to find the instantons, as well as the DNS and the computation of the regularized Fredholm determinant have been formulated and implemented in real variables, i.e.~$\psi = \psi_r + \mathbbm{i} \psi_i$ and likewise for all other fields.  
\par 
The NLS~\eqref{eq:nls} in this formulation reads: 
\begin{align}
    \partial_t 
    \begin{pmatrix}
        \psi_r \\ \psi_i
    \end{pmatrix}
    &= 
    \begin{pmatrix}
        - \nu \partial_x^4 & -\frac{1}{2} \partial_x^2 \\  \frac{1}{2} \partial_x^2 & - \nu \partial_x^4
    \end{pmatrix}
    \begin{pmatrix}
        \psi_r \\ \psi_i
    \end{pmatrix}
    + \begin{pmatrix}
     -\left(\psi_r^2 + \psi_i^2\right)^3 \psi_i \\[0.4em]
    \phantom{-}\left(\psi_r^2 + \psi_i^2\right)^3 \psi_r 
    \end{pmatrix}
    + \begin{pmatrix}
        \chi^{1/2} * \eta_r \\ \chi^{1/2} * \eta_i 
    \end{pmatrix}
    \,. 
\end{align}
For the DNS, since we work in a real formulation, we have to enforce the correct symmetries for the noise variable when sampling the noise term in Fourier space. More on the implementation of the Gaussian forcing and the symmetries can be found in~\cite{lang-2011,schorlepp-2022-nse-supp}.
The adjoint PDE~\eqref{eq:adjoint pde} reads: 
\begin{align}
    &\partial_t 
    \begin{pmatrix}
    z_r \\
    z_i
    \end{pmatrix}
    = \begin{pmatrix}
    \nu \partial_x^4 &  -\frac{1}{2}\partial_x^2  \\
    \frac{1}{2}\partial_x^2  & \nu \partial_x^4
    \end{pmatrix}
    \begin{pmatrix}
    z_r \\ 
    z_i
    \end{pmatrix}
    +
    \begin{pmatrix}
    6 \psi_r \psi_i \lvert\psi\rvert^{4} & 
    -6 \psi_r^2 \lvert\psi\rvert^{4}  - \lvert\psi\rvert^{6}
     \\
      6 \psi_i^2 \lvert\psi\rvert^{4}  + \lvert\psi\rvert^{6}
     & -6 \psi_r \psi_i \lvert\psi\rvert^{4}
    \end{pmatrix}
     \begin{pmatrix}
    z_r \\
    z_i
    \end{pmatrix}
    \, .
\end{align}
The regularized Fredholm determinant~$\mathrm{det}'(\mathrm{Id} - B_a)$ is approximated using the largest eigenvalues~$\kappa_a^{(i)} \neq 1$ of  the  operator~$B_a$ given in equation~\eqref{eq:definition integral operator fredholm determinant appendix} with the projection operator~$\operatorname{pr}_{\eta_a^{\perp}}$ defined in equation~\eqref{eq:definition projection operator fredholm determinant}. 
The  eigenvalues of~$B_a$ are calculated iteratively by applying this operator to test vectors~\cite{lehoucq-1998}: By again employing  the adjoint-state method, one can show that the second variation operator acts on a fluctuation~$\delta \eta$ in the following way~\cite{schorlepp-2023b}: 
\begin{align}
    \left. \frac{\delta^2 (\lambda_aF)}{\delta \eta^2}\right\rvert_{\eta = \eta_a^\varphi}
     \delta \eta 
    = \chi^{1/2}*\zeta \,, 
\end{align}
where~$\zeta$ is found from the solution of second-order equations:  
\begin{equation}
\begin{aligned}\label{eq:second variation}
    &\partial_t
    \begin{pmatrix}
    \gamma_r \\ \gamma_i
    \end{pmatrix}
    = \begin{pmatrix}
    -\nu \partial_x^4 &  -\frac{1}{2}\partial_x^2  \\[0.4em]
    \frac{1}{2}\partial_x^2  & -\nu \partial_x^4
    \end{pmatrix}
     \begin{pmatrix}
    \gamma_r \\ \gamma_i
    \end{pmatrix}
    + 
     \begin{pmatrix}
    -6 \psi_r \psi_i \lvert\psi\rvert^{4}  & 
     -6 \psi_i^2 \lvert\psi\rvert^{4}  - \lvert\psi\rvert^{6}
     \\[0.4em]
     6 \psi_r^2 \lvert\psi\rvert^{4}  + \lvert\psi\rvert^{6} & 6 \psi_r \psi_i \lvert\psi\rvert^{4}
    \end{pmatrix}
     \begin{pmatrix}
    \gamma_r \\ \gamma_i
    \end{pmatrix}
    + \begin{pmatrix}
        \chi^{1/2} * \delta \eta_r \\ \chi^{1/2} * \delta  \eta_i 
    \end{pmatrix}
    \\[0.4em]
     &\partial_t 
    \begin{pmatrix}
    \zeta_r \\  \zeta_i
    \end{pmatrix}
    = \begin{pmatrix}
    \nu \partial_x^4 &  -\frac{1}{2}\partial_x^2  \\
    \frac{1}{2}\partial_x^2  & \nu \partial_x^4
    \end{pmatrix}
    \begin{pmatrix}
    \zeta_r \\  \zeta_i
    \end{pmatrix}
    +
    \begin{pmatrix}
    6 \psi_r \psi_i \lvert\psi\rvert^{4} & 
    -6 \psi_r^2 \lvert\psi\rvert^{4}  - \lvert\psi\rvert^{6}
     \\
      6 \psi_i^2 \lvert\psi\rvert^{4}  + \lvert\psi\rvert^{6}
     & -6 \psi_r \psi_i \lvert\psi\rvert^{4}
    \end{pmatrix}
     \begin{pmatrix}
    \zeta_r \\  \zeta_i
    \end{pmatrix}
    + A_1 z_r 
    \begin{pmatrix}
    \gamma_r \\ \gamma_i
    \end{pmatrix}
    + A_2 z_i 
    \begin{pmatrix}
    \gamma_r \\ \gamma_i
    \end{pmatrix}
    \,, 
\end{aligned}
\end{equation}
with 
\begin{align}
    &\begin{pmatrix}
    \gamma_r(\cdot, 0) \\ \gamma_i(\cdot, 0)
    \end{pmatrix}
    = 0
    \,, \quad 
    \begin{pmatrix}
    \zeta_r(x, T) \\  \zeta_i(x, T)
    \end{pmatrix} 
    =  4 \nu   \lambda_a \,  \delta''(x)
    \begin{pmatrix}
       \partial_x^2 \gamma_r(0,T)
        \\
        \partial_x^2 \gamma_i(0,T)
    \end{pmatrix}
    \,, 
    \\[0.4em] 
    &A_1 
    = 6\lvert\psi\rvert^2
    \begin{pmatrix}
     \psi_i  \left(\lvert\psi\rvert^2 + 4 \psi_r^2 \right)
    &    \psi_r  \left(\lvert\psi\rvert^2 + 4 \psi_i^2 \right)
     \\
      \psi_r  \left(\lvert\psi\rvert^2 + 4 \psi_i^2 \right)  &  \psi_i  \left(3 \lvert\psi\rvert^2 + 4 \psi_i^2   \right)
    \end{pmatrix}
    \,, \quad 
    A_2 
    = -6\lvert\psi\rvert^2
    \begin{pmatrix}
     \psi_r  \left(3\lvert\psi\rvert^2 + 4 \psi_r^2 \right)
    &    \psi_i  \left(\lvert\psi\rvert^2 + 4 \psi_r^2 \right)
     \\
      \psi_i  \left(\lvert\psi\rvert^2 + 4 \psi_r^2 \right)  &  \psi_r  \left(\lvert\psi\rvert^2 + 4 \psi_i^2   \right)
    \end{pmatrix}
    \,,  
\end{align}
where $\psi = \psi_a^\varphi, z= z_a^\varphi$ are background fields found from the preceding  instanton computation.  
\setcounter{equation}{0}
\renewcommand{\theequation}{D.\arabic{equation}}
\section{Appendix D: Analytical solution of the linear part of the NLS instantons}
In this appendix, we analytically  solve the instanton equations and compute the prefactor 
in case of vanishing nonlinearity for an arbitrary observable value~$a$ of the energy dissipation density. For this, we use that at the instanton, $\eta_a = \chi^{1/2}*z_a$ and $\varepsilon(0,T) =a$ in equation~\eqref{eq:nls} and~\eqref{eq:adjoint pde} and we Fourier transform in space, such that the following system is considered: 
\begin{align} 
    &\partial_{t} \widehat{\psi_a} + \frac{\mathbbm{i}}{2} \Tilde{k}^2 \widehat{\psi_a} +\nu \Tilde{k}^4 \widehat{\psi_a}  =  \ell \widehat{\chi} \, \widehat{z_a} \,,\quad  \psi_a(\cdot,t=0) = 0 \,, 
    \label{eq:nls linear instanton}
    \\
     &\partial_t \widehat{z_a} +  \frac{\mathbbm{i}}{2} \Tilde{k}^2 \widehat{z_a} - \nu \Tilde{k}^4 \widehat{z_a}  = 0 \,, \qquad 
    \widehat{z_a}(\Tilde{k}, T)  = - 
    \frac{\lambda_a}{\ell}  \, 
    \Tilde{k}^2 \beta_a
   \, .
\end{align}
where $\Tilde{k} = \frac{2 \pi k}{\ell}$ for $k \, \in \, \mathbb{Z}$ and $\beta_a = 4 \nu \,\partial_x^2 \psi_a(0,T)$. Note that the complex phase of~$\beta_a$ is arbitrary, which gives rise to the zero mode, but only~$\lvert \beta_a\rvert^2$ enters the instanton action below.  
The solution for the adjoint field reads
\begin{align}
    \widehat{z_a}(\Tilde{k},t) = 
    \frac{-\lambda_a \beta_a \Tilde{k}^2}{\ell} \exp\left(-\frac{\mathbbm{i}}{2}\Tilde{k}^2 (t-T) +\nu \Tilde{k}^4 (t-T)\right) \,, 
    \label{eq:solution adjoint field linear}
\end{align}
with $ \widehat{\chi}(k)  = \frac{1}{\ell}  \mathbbm{1}_{\left( 0 < \lvert k \rvert  < 0.3 \frac{\ell}{2 \pi}\right)}$ for $k \in \mathbb{Z}$, as in the main text. 
The instanton action reads:
\begin{align}
    S_{\mathrm{I}}(a) &= \frac{1}{2} \ell^2 \int_{0}^T  \sum_{k \, \in \, \mathbb{Z}} \left\lvert\widehat{z_{\mathrm{a}}}\left(\frac{2\pi}{\ell}k,t\right)\right\rvert^2  \,  \widehat{\chi}(k)  \, \mathrm{d} t
    \, .
\end{align}
Substituting $\widehat{\chi}$, the solution for~$\widehat{z_a}$ and $\lvert \beta_a\rvert^2 = 8\nu a$, and using the fact that the expression does not depend on the sign of $k$, we obtain:
\begin{align}
      S_{\mathrm{I}}(a) &= 
    \frac{4\lambda_a^2 a}{\ell}  \sum_{k \, \in \, (0, \,  0.3\frac{\ell}{2 \pi} ) \, \cap \, \mathbb{N}}  \,
    \left[ 1- \exp\left(-2 \nu \left(\frac{2 \pi k}{\ell} \right)^4 T\right) \right]
    \,.   
    \label{eq:instanton action Fourier}
\end{align}
Now, to find the relation between the action and the observable value $a$, we need to obtain the explicit dependence~$\lambda_a$ in~$a$. For this, we solve equation~\eqref{eq:nls linear instanton}. Its solution is: 
\begin{align}
   \widehat{\psi_a}(\Tilde{k},t) =  -\mathbbm{1}_{(\lvert \tilde{k}\rvert \, \in \, (0, \,  0.3 ))}  \, \exp\left(-\frac{\mathbbm{i}}{2}\Tilde{k}^2 (t-T) -\nu \Tilde{k}^4 (t-T)\right) \,  \frac{\lambda_a \beta_a}{2 \nu \ell \tilde{k}^2}
    \left(\exp\left(2 \nu \tilde{k}^4 (t-T)\right) - \exp\left(-2 \nu \tilde{k}^4 T\right)\right)
\end{align}
We use the final time condition: $\varepsilon(0,T) = 2 \nu \lvert\partial_x^2 \psi(0,T)\rvert^2  = a$ and compute the inverse Fourier transform of the above solution. The Lagrange multiplier is then
\begin{align}
    \lambda_a^2 
    = \frac{\ell^2}{16} \left(\sum_{k \, \in \, (0, \,  0.3\frac{\ell}{2 \pi} ) \, \cap \, \mathbb{N}} \, \left[1-  \exp\left(-2 \nu \left(\frac{2 \pi k}{\ell} \right)^4 T\right)  \right] \right)^{-2}
\end{align}
Substituting in equation~\eqref{eq:instanton action Fourier},  the instanton action for the linear part of the NLS becomes:
\begin{align}
    S_{\mathrm{I}}(a) 
    =  \frac{a\ell}{4}  
    \left(\sum_{k \, \in \, (0, \,  0.3\frac{\ell}{2 \pi} ) \, \cap \, \mathbb{N}}   \,  \left[1-  \exp\left(-2 \nu \left(\frac{2 \pi k}{\ell} \right)^4 T\right)  \right] \right)^{-1}
    = c a \,, 
\end{align}
i.e.~the instanton action in this case is a linear function in the observable value~$a$ with slope~$c$. This gives the exponential contribution to the PDF in equation~\eqref{eq:instanton pdf subexponential prefactor}. 
As argued below, the prefactor~$C(a)$ does not depend on~$a$ here and is therefore fixed by normalization. Hence, the energy dissipation density is exponentially distributed: 
\begin{align}
    \rho(a) = \frac{c}{\sigma^2} \exp\left(-\frac{c}{\sigma^2}a \right) \, .
\end{align}
In equation~\eqref{eq:prefactor formula}, $\lambda_a = \mathrm{d}S_{\mathrm{I}}/\mathrm{d}a = c$ does not depend on~$a$, since the instanton action is linear in~$a$. The Fredholm determinant $\mathrm{det}'(\mathrm{Id} - B_a)$ does not depend on~$a$ either, which can be seen upon inspecting the projection operator in equation~\eqref{eq:definition projection operator fredholm determinant} with $\eta_a^\varphi = \chi^{1/2}* z_a$ with~$z_a$ from equation~\eqref{eq:solution adjoint field linear}, as well as the second variation~\eqref{eq:second variation} in case of vanishing nonlinearity.  
\end{document}